\documentclass[10pt,twocolumn,longbibliography,amsmath,amssymb,floatfix,superscriptaddress,aps,prl]{revtex4-2}

\usepackage{graphicx}
\usepackage{dcolumn}
\usepackage{bm}
\usepackage{color}
\usepackage{txfonts}
 \usepackage{amsmath}
 \usepackage{hyperref}

 \begin{abstract}

Nonlinear Compton scattering driven by ultraintense lasers presents a promising avenue for enhancing the photon energy, brilliance, and setup compactness of $\gamma$-ray sources. However, a significant challenge lies in achieving a high polarization degree with commonly generated unpolarized electrons, thus addressing a longstanding puzzle in the field. Here we investigate the polarization dynamics of photons emitted by an unpolarized electron beam interacting with a counter-propagating ultraintense laser pulse numerically, and propose a novel method to generate highly polarized $\gamma$ rays via nonlinear Compton scattering with the aid of vacuum dichroism effect. Our simulations reveal that high-brilliance $\gamma$ rays with polarization beyond 90\% are feasible in a single-shot interaction, rivaling the highest achieved by any $\gamma$-ray sources to date, based on a developed Monte Carlo method incorporating polarization-resolved tree processes of nonlinear Compton scattering and Breit-Wheeler pair production and one-loop vacuum polarization. This generation method showcases an extraordinary ultra-high polarization degree and a user-friendly all-optical experimental setup, while harnessing the high photon energy and brilliance characteristic of nonlinear Compton scattering sources, thus making it of great potential for experimental applications.
\end{abstract}

\begin{document}
\title{Achieving High Polarization of Photons  Emitted by Unpolarized Electrons in Ultrastrong Laser Fields}
\author{Xian-Zhang Wu}
\affiliation{Department of Nuclear Science and Technology, Xi'an Jiaotong University, Xi'an 710049, China}
\author{Yan-Fei Li}\email{liyanfei@xjtu.edu.cn}    
\affiliation{Department of Nuclear Science and Technology, Xi'an Jiaotong University, Xi'an 710049, China}
\author{Yu-Tong Li}\email{ytli@iphy.ac.cn}  
\affiliation{Beijing National Laboratory for 
Condensed Matter Physics, Institute of Physics, CAS, Beijing 100190,
China}
\affiliation{School of Physical Sciences, University of Chinese Academy of Sciences, Beijing 100049, China}
\affiliation{Collaborative Innovation Center of IFSA (CICIFSA), Shanghai Jiao Tong University, Shanghai 200240, China}
\affiliation{Songshan Lake Materials Laboratory, Dongguan, Guangdong 523808, China}
\date{\today}
\maketitle

With the widespread applications of high-energy $\gamma$ rays across various scientific domains, including astrophysics, particle physics, nuclear physics, and high-energy physics \cite{Fleck2020,BADELEK2004,Corde2013,Howell_2022}, as well as in medical imaging \cite{Glinec2005,Sakdinawat2010} and national security \cite{Albert_2014}, there has been a global effort to construct facilities capable of generating high-intensity $\gamma$ rays. Examples of such facilities include FACET-II at SLAC National Accelerator Laboratory \cite{FACETII2022}, ELI-NP (Extreme Light Infrastructure - Nuclear Physics) \cite{Tanaka2020ELINP}, SPring-8 \cite{SPring8}, and HT$\gamma$S (High-Intensity $\gamma$-ray Source) at Duke University \cite{HIGSDuke}. Photon polarization, as a fundamental parameter for source quality, plays a crucial role in elucidating information about the parity and spin of nuclear states \cite{Schaerf2005,Speth_1981}, the crystal structure of matter \cite{Grant1972}, predictions in quantum electrodynamics (QED) \cite{Piazza2012,King2016, Ilderton2016,Bragin2017,Liyf2019,Hu2020}, and various other aspects \cite{Akbar2017,Moortgat2008,Uggerh2005}.  As such, the purification of photon polarization in $\gamma$-ray sources is a longstanding challenge critical for leveraging the potential of polarized $\gamma$ rays in scientific research and practical applications.

Polarized high-energy $\gamma$ rays are primarily generated through Bremsstrahlung or Compton scattering processes.  
The $\gamma$ rays emitted from Bremsstrahlung naturally exhibit linear polarization when photons are collimated within a small solid angle away from the direction of the primary electrons, with a maximum degree typically on the order of 30-50\% \cite{Cabibbo1962,Heitler1954}. However, this polarization diminishes significantly due to depolarization effects caused by multiple scatterings in thick targets designed for higher yield. Nevertheless, post-selection techniques can enhance the polarization degree to approximately 80\% by utilizing coherent radiation from electrons interacting with the periodic potential of a crystal, particularly when the crystal lattice aligns relative to the electron beam \cite{LOHMANN1994,Rambo1998}. Circular polarization can be attained either through helicity transfer from longitudinally polarized electrons \cite{McVoy1957,APYAN1998} or via birefringence \cite{KIRSEBOM1999, APYAN2005}, a phenomenon that converts the linear polarization of the photon beam into circular polarization. Overall,  incoherent Bremsstrahlung struggles to achieve high linear polarization and photon yield [or brilliance, typically $\lesssim$ $ 10^{18}$ phs/(s mm$^2$ mrad$^2$ 0.1\%BW) \cite{Giulietti2008,Sarri2014}] is limited when suppressing depolarization caused by the unavoidable multiple scattering in the Coulomb field of nuclei (known as Mott scattering) \cite{Baier1998,POTYLITSIN1997,McMaster1961}. Conversely, coherent Bremsstrahlung encounters limitations in terms of higher polarization ($\gtrsim 90\%$) and radiation flux primarily due to the damage threshold of the crystal material and the strict requirements for alignment operations \cite{Uggerh2005, Biryukov1997,timm1969coherent}.

Compton scattering $\gamma$-ray sources operating in linear regimes have the capacity to achieve higher polarization. They acquire the polarization characteristics of incident laser photons, with the polarization degree increasing as the emitted photon energy rises \cite{Federici1980,Berestetskii1982,NAKANO1998,BOCQUET1997}. Notably, the polarization degree reaches approximately 100\% at the Compton edge for both linear and circular polarization, gradually decreasing with decreasing energy until it eventually reaches zero. Linear Compton scattering sources face inherent limitations in providing significant photon energy, primarily due to the fundamental constraint expressed by $\omega_\gamma\lesssim 4\gamma^2 \omega_0$, where $\gamma$ represents the electron Lorenz factor and $\omega_0$ ($\omega_\gamma$) denotes the energy of the laser photon (emitted $\gamma$ ray). Additionally, the photon yield 
of $N_\gamma \ll N_e$, is constrained by the low collision luminosity, where $N_e$ is the number of the seed electrons. The remedy for these limitations involves utilizing high-intensity lasers, in which the interaction regime moves into the nonlinear domain. The sparked increased interest in nonlinear Compton scattering is further propelled by the rapid development of ultra-intense laser technology with a record peak intensity of $I_0 \approx 10^{23} \mathrm{W/cm^2}$ \cite{Yoon2021,Danson2019} facilitated by advancements such as Chirped Pulse Amplification (CPA) technology \cite{STRICKLAND1985}. 
Given that both the ratio of typical emitted photon energy to electron energy $\delta=\omega_\gamma/\varepsilon_e$ and the photon number $N_\gamma$ are proportional to $\sqrt{I_0}/\omega_0$ \cite{Piazza2012}, strong-nonlinear Compton scattering driven by ultra-intense lasers is recognized for its capability to produce high-energy $\gamma$ rays with exceptionally high brilliance \cite{Sarri2014,Albert_2016,Zhu2018,King2016,Liyf2020,King2013}. However, the polarization degree decreases dramatically with laser intensity \cite{King2016,King2013,Liyf2020,Tang2024}, as the radiation formation length is much smaller than the laser wavelength, resulting in the laser field remaining relatively constant during photon formation, resembling the incoherent Bremsstrahlung emission process. The necessity of initial polarized electrons for generating highly polarized $\gamma$ rays in nonlinear Compton scattering is acknowledged \cite{Liyf2020, Gonoskov2022}. This requirement  poses significant challenges to experiments, as electron beams produced via laser-plasma wakefield acceleration (LWFA) and traditional accelerators are typically unpolarized \cite{Gonoskov2022,Esarey2009,Moortgat2008}. This challenge is especially pronounced in potential all-optical setups, where the generation of polarized electrons has not yet been experimentally demonstrated \cite{Gonoskov2022,Wen2019}. Thus, the question arises: can high polarization of photons emitted by unpolarized electrons in ultra-intense laser fields be achieved?

\begin{figure}[t]
    \centering
    \includegraphics[width=1\linewidth]{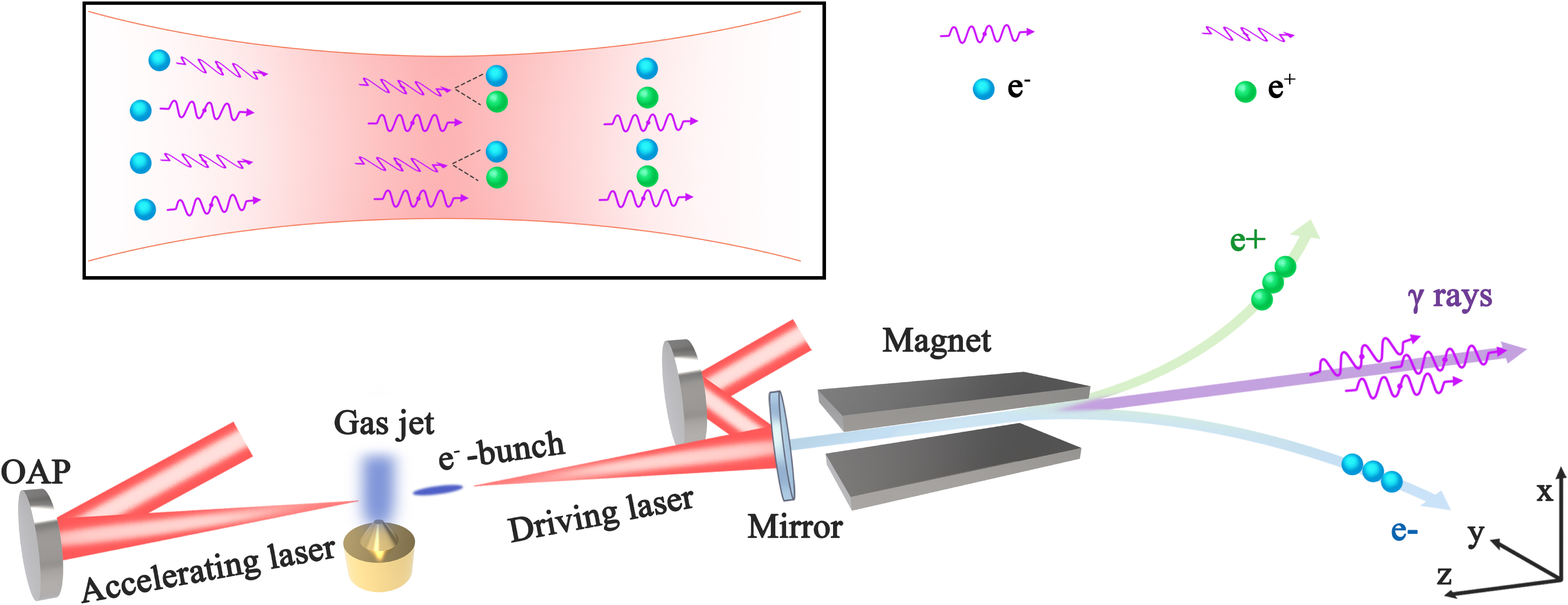}
    \caption{Scenario for generating highly polarized $\gamma$ rays via ultraintense-laser-electron interactions. A linearly polarized laser pulse (polarization along the $x$-axis) propagates along the $+z$ direction and collides head-on with an unpolarized electron beam from the laser wakefield accelerator, resulting in the generation of linearly polarized $\gamma$ rays (also polarized along the $x$-axis). The polarization degree is significantly enhanced by the vacuum dichroism effect, corresponding to the asymmetric pair production between $\xi_3=+1$ and $\xi_3=-1$ photons as they propagate in the laser field, as shown in the inset. A magnet is utilized to isolate the desired $\gamma$ rays from electrons and positrons. }
    \label{fig1}
    
    \begin{picture}(300,0)
        \put(164,226){\tiny{$\xi_{3}=+1$}}
        \put(204,226){\tiny{$\xi_{3}=-1$}}
    \end{picture}
\end{figure}

In this letter, we investigate the feasibility of generating highly linearly polarized brilliant GeV-$\gamma$ rays via nonlinear Compton scattering with unpolarized electrons (see Fig.\ref{fig1}). We show numerically that although the photon polarization is initially around 0\% in the high-energy edge of nonlinear Compton scattering as well known, surprisingly, it can be significantly improved to over 90\% in a single-shot interaction with the assistance of the vacuum dichroism effect, which involves asymmetric photon decay (pair production) between two opposite polarization states. As a result, the photons exhibit a high polarization along the direction of polarization that experiences less decay. Furthermore, this high linear polarization can be converted into circular polarization, either through the crystal birefringent effect \cite{Adamczyk2003,Uggerh2005} or via vacuum birefringence using ultra-intense lasers \cite{Lv2024,Dai2024}.

This investigation is focused on accurately modeling the polarization dynamics of emitted photons in ultrastrong electromagnetic fields. To achieve this, we have developed the Monte Carlo simulation method \cite{Liyf2020,Liyfei2020,Liyf2022,Liyf2023}, with the capability to effectively handle the intricate photon polarization dynamics induced by the strong-field vacuum polarization particularly including vacuum birefringence (VB) and vacuum dichroism (VD), manifestations of QED in extreme conditions, where the photon dispersion relation is altered by the background field via radiative corrections induced by virtual particles\cite{Berestetskii1982,Bragin2017}. The polarization of a photon emitted by an electron is determined by the spin- and polarization-resolved radiation probability, as detailed in the methodology of our previous publication \cite{Liyf2020}. As the photon propagates within the laser field, its evolution is influenced by VB and VD. The VB, akin to optical birefringence, splits light into two orthogonal polarizations with distinct refractive indices, characterized by the rotation of Stokes parameters [${\bm \xi}=(\xi_1,\xi_2,\xi_3)$] between $\xi_1$ and $\xi_2$ \cite{Bragin2017,Dinu2014}. In contrast, the VD, which denotes the differential absorption or transmission of light by the vacuum based on its polarization state, is achieved through a no-pair-production-evolution in each time step of $\Delta t$ \cite{Zhuang2023,CAIN,SM}:
\begin{widetext}
\begin{equation}\label{xi}
    {\bm \xi}_f=\frac{{\bm \xi}_i \left\{1-\int_0^1 dx C_p \left[{\rm IntK}_{\frac{1}{3}}(\rho)+(\frac{x}{1-x}+\frac{1-x}{x}){\rm K}_{\frac{2}{3}}(\rho)\right]\Delta t\right \}+\int_0^1 dx C_p  {\bf \hat e}'_3{\rm K}_{\frac{2}{3}}(\rho) \Delta t}{1-\int_0^1 dx C_p \left[ (\frac{x}{1-x}+\frac{1-x}{x}){\rm K}_{\frac{2}{3}}(\rho)+{\rm IntK}_{\frac{1}{3}}(\rho)-\xi_3{\rm K}_{\frac{2}{3}}(\rho)\right]\Delta t},
\end{equation}
\end{widetext}
where $C_p=\alpha m^2/(\sqrt{3}\pi\omega_\gamma)$ with $\alpha\approx1/137$ the fine structure constant and $m$ the electron mass; $\rho=2/[3\chi_\gamma x(1-x)]$; ${\bf \hat e}'_3=(0,0,1)$; ${\rm IntK}_{\frac{1}{3}}(\rho)\equiv \int_{\rho}^{\infty} {\rm d}z {\rm K}_{\frac{1}{3}}(z)$,  with ${\rm K}_n$ being the $n$-order modified Bessel function of the second kind;  $\omega_\gamma$ is the photon energy; and ${\boldsymbol \xi}_{i,f} \leq 1$ refers to the photon polarization vector before and after a time step with  the Stokes parameters defined with respect to the axes of ${\bf \hat e}_1$ and ${\bf \hat e}_2$ \cite{McMaster1961}; $\hat{{\bf e}}_1$ is the unit vector along the direction of the transverse component of electron acceleration, $\hat{{\bf e}}_2=\hat{\bf e}_v\times\hat{\bf e}_1$ with $\hat{\bf e}_v$ the unit vector along photon momentum. The $\xi_3=+1$ ($\xi_3=-1$) represents linear polarization along ${\bf \hat e}_1$ (${\bf \hat e}_2$) axis, respectively.  Quantum parameter is defined as $\chi_{\gamma,e} \equiv |e| \sqrt{-(F_{\mu v}k^v)^2}/m^3$ with $F_{\mu\nu}$ being the field tensor and $k^v$ the four-vector of photon (or electron) momentum. Relativistic units $\hbar=c=1$ are used throughout. Note that electron (positron) spin and photon polarization are described in mixed states in the simulations.

The no-pair-production-evolution of photon polarization similar to the no-emission-spin-variation of electrons \cite{Liyf2023,Liyf2022,Torgrimsson_2021}, originates from the dependence of 
 pair-production probability  on photon polarization  \cite{Baier1998,Liyfei2020}: 
\begin{equation}\label{Wpair}
\frac{{\rm d^2}W_{\rm pair}}{{\rm d}x {\rm d}t}=C_P\left[(\frac{x}{1-x}+\frac{1-x}{x}){\rm K}_{\frac{2}{3}}(\rho)+{\rm IntK}_{\frac{1}{3}}(\rho)-\xi_3{\rm K}_{\frac{2}{3}}(\rho)\right],
\end{equation}
where $x\in(0,1)$ corresponds to the ratio of newborn positron energy to its parent photon energy. The process where a high-energy photon decays into an electron-positron pair is determined by Eq.(\ref{Wpair}). For a more comprehensive description of our methodology, including detailed algorithms and formulas, please refer to the Supplemental Material \cite{SM}.

\begin{figure}[t]
    \includegraphics[width=1\linewidth]{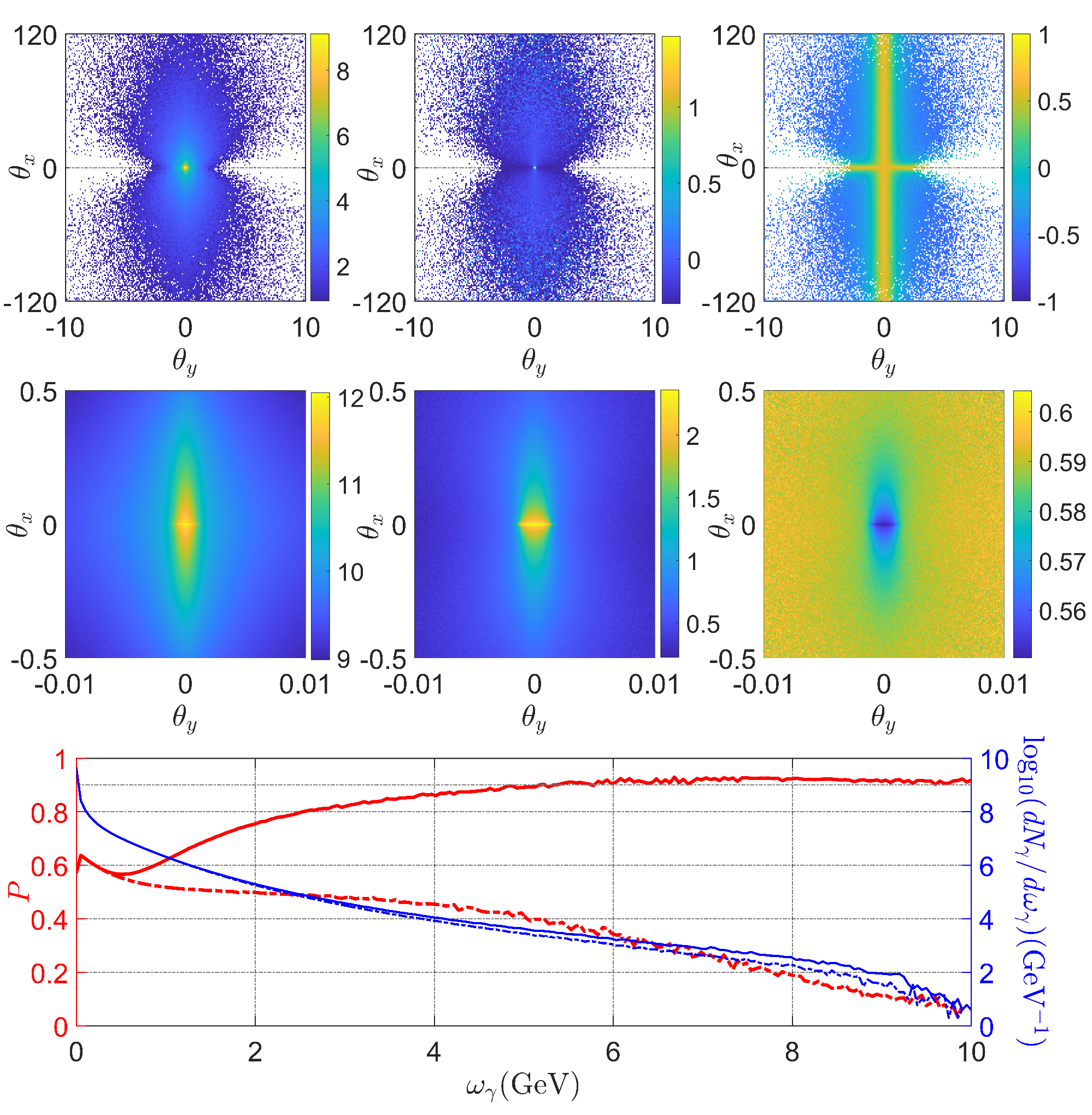}
     \begin{picture}(300,0)
  \put(15,248){(a)}

  \put(95,248){(b)}

   \put(172,248){(c)}

     \put(15,165){\color{white}(d)}

  \put(95,165){\color{white}(e)}

   \put(172,165){\color{white}(f)}

\put(20,85){(g)}
\end{picture}	
    \caption{(a)-(c): Distributions of number density log$_{10}$($\mathcal{N}_\gamma$), average photon energy log$_{10}(\overline{\omega}_\gamma$) (MeV), and $\overline{\xi}_3$, with respect to scattering angles of $\theta_x\equiv k_x/k_z$ and $\theta_y\equiv k_y/k_z$, for the whole emitted photons. Here, $\mathcal{N}_\gamma=$ d$^2N_\gamma/$d$\theta_x$d$\theta_y$. (d)-(f): Distributions of log$_{10}$($\mathcal{N}_\gamma$), log$_{10}(\overline{\omega}_\gamma)$ (MeV), and $\overline{\xi}_3$, for the selected photons with $\theta_x 
    \in [-0.5, 0.5]$ and $\theta_y
    \in [-0.01, 0.01]$. (g): Polarization of $P=\overline{\xi}_3$ (red), and log$_{10}($d${N_\gamma}/$d$\omega_\gamma$) (blue) for the selected photons in (d) vs  photon energy of $\omega_\gamma$. The solid and dash-dotted lines indicate the results with vacuum polarization effect included and excluded, respectively.}
    \label{fig2}
\end{figure}
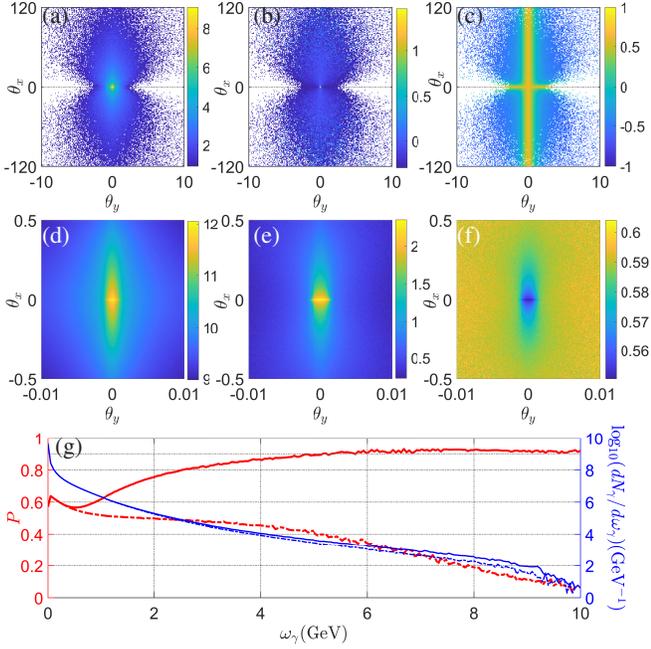

  A typical simulation result for production of highly polarized $\gamma$ rays with a realistic tightly-focused Gaussian laser pulse \cite{Yousef2002} is shown in Fig.\ref{fig2}.  The peak laser intensity is $I_0\approx 3.09\times 10^{22}$ W/cm$^2$ $ (a_0=150)$, pulse duration (the full width at half maximum, FWHM) 
   $\tau=10T_0$ with $T_0$ the period, wavelength $\lambda=1 \mu$m, and focal radius $w_0=5 \lambda$, and the polarization is linear along $x$ direction. The colliding electron bunch is set with features of laser-accelerated electron sources \cite{Esarey2009,Gonsalves2019,Leemans2014} for a potential favorable all-optical setups. $N_e=1\times 10^6$ electrons uniformly distributed longitudinally and normally distributed transversely in a cylindrical form at length of $L_e=3 \lambda$ and standard deviation of $\sigma_{x,y}=0.3 \lambda$. The initial mean kinetic energy is $\varepsilon_e=10$ GeV, the energy spread $\Delta\varepsilon_e/\varepsilon_e=10\%$ (FWHM), the angular divergence $\Delta \theta=1$ mrad (FWHM) and the electron spin-polarization vector ${\bf S}_i=(0,0,0)$. The laser intensity and electron energy are chosen above to keep $\chi_e^{\rm max}\approx5\times10^{-6} a_0 \gamma \approx 14.67$ for substantial high-energy photon emission and VD influence, consequently accompanied by a nontrivial pair production. However, the influence of photons emitted by newly created pairs on the final $\gamma$-ray polarization in the desired high-energy range has been demonstrated to be negligible \cite{SM}.
   
The scattering angular distributions of the total produced photons exhibit a wide range, varying from -120 to 120 in $\theta_x$ and from -10 to 10 in $\theta_y$. The number density, average photon energy and polarization of $\overline{\xi}_3$ ($\overline{\xi}_1\approx 0$ and $\overline{\xi}_2\approx 0$ \cite{SM}) decreases gradually from the center to the edges, as shown in Figs. \ref{fig2}(a)-(c). The total photon yield with energies of $\omega_\gamma>1$ MeV amounts to $1.92\times10^8$ ($192 \times N_e$), with a duration determined by the electron bunch length: $\tau_\gamma\approx \tau \approx 10$ fs, an average photon energy of 48.19 MeV and an average polarization of 62.51\%. 
For the application of a collimated high-density beam, photons within $\theta_x\in(-0.5,0.5)$ and $\theta_y\in(-0.01,0.01)$ are selected for a more detailed analysis, as illustrated in Figs. \ref{fig2}(d)-(g). This selection results in a photon yield of $1.58\times10^8$ with energies of $\omega_\gamma>1$ MeV, a flux of $1.58\times10^{22}$ s$^{-1}$ and a polarization of 60.98\%. The photon polarization in the center of the angular distribution is slightly lower than that in the surrounding areas due to the relatively larger photon energy [see Fig. \ref{fig2}(e)], corresponding to the $\omega_\gamma\in$ (50 MeV, 600 MeV) range in Fig. \ref{fig2}(g) where photon polarization decreases with energy. 

The polarization shows an energy-dependence feature as illustrated in Fig. \ref{fig2}(g), where the photon polarization is directly proportional to the $\gamma$-photon energy when $\omega_\gamma \ge 0.6$ GeV, reaching a maximum of 92\% when $\omega_\gamma$ is close to 6.5 GeV and keeping stable until $\omega_\gamma \approx 10$ GeV. Purification of photon polarization through post-energy selection, akin to other polarized $\gamma$-ray sources, is admired for applications. For instance, for $\gamma$ photons with energy higher than 2 GeV, 4GeV, 6GeV and 8GeV, polarization degrees are 79.92\%, 88.85\%, 91.74\% and 91.99\%. For a feasible 10 pC charge of a laser-driven electron bunch \cite{Gonsalves2019}, the brilliance will be $1.48\times 10^{22}$, $5.23\times 10^{21}$, $5.18\times 10^{21}$ and $1.64\times 10^{21}$ phs/(s mm$^2$ mrad$^2$ 0.1\%BW) at 2 GeV, 4GeV, 6GeV and 8GeV, respectively.

We also conducted simulations excluding vacuum polarization artificially in Fig.\ref{fig2}(g). The photon polarization decreases with $\omega_\gamma$ to zero when $\omega_\gamma \approx 10$ GeV, consistent with the findings in other studies \cite{King2016,Liyf2020,Tang2024}. Therefore, the high polarization of high-energy photons could also serve as a distinct signature for observing vacuum polarization, a fundamental prediction of quantum electrodynamics (QED) \cite{Berestetskii1982}.

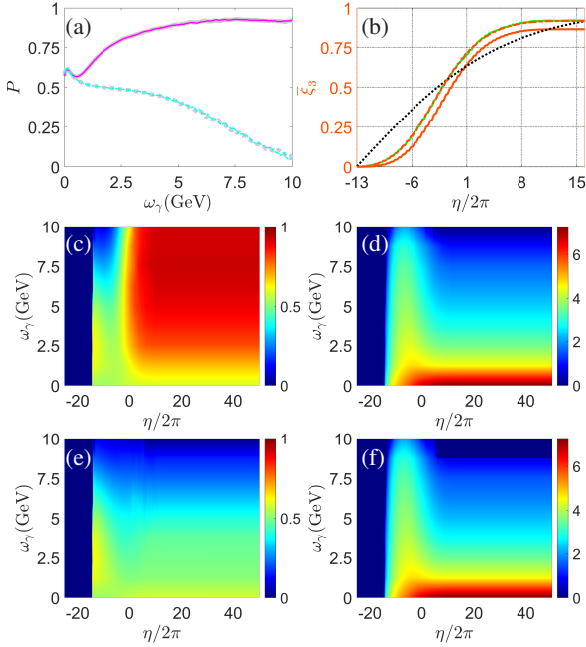
\begin{figure}[tbh]
    \includegraphics[width=1\linewidth]{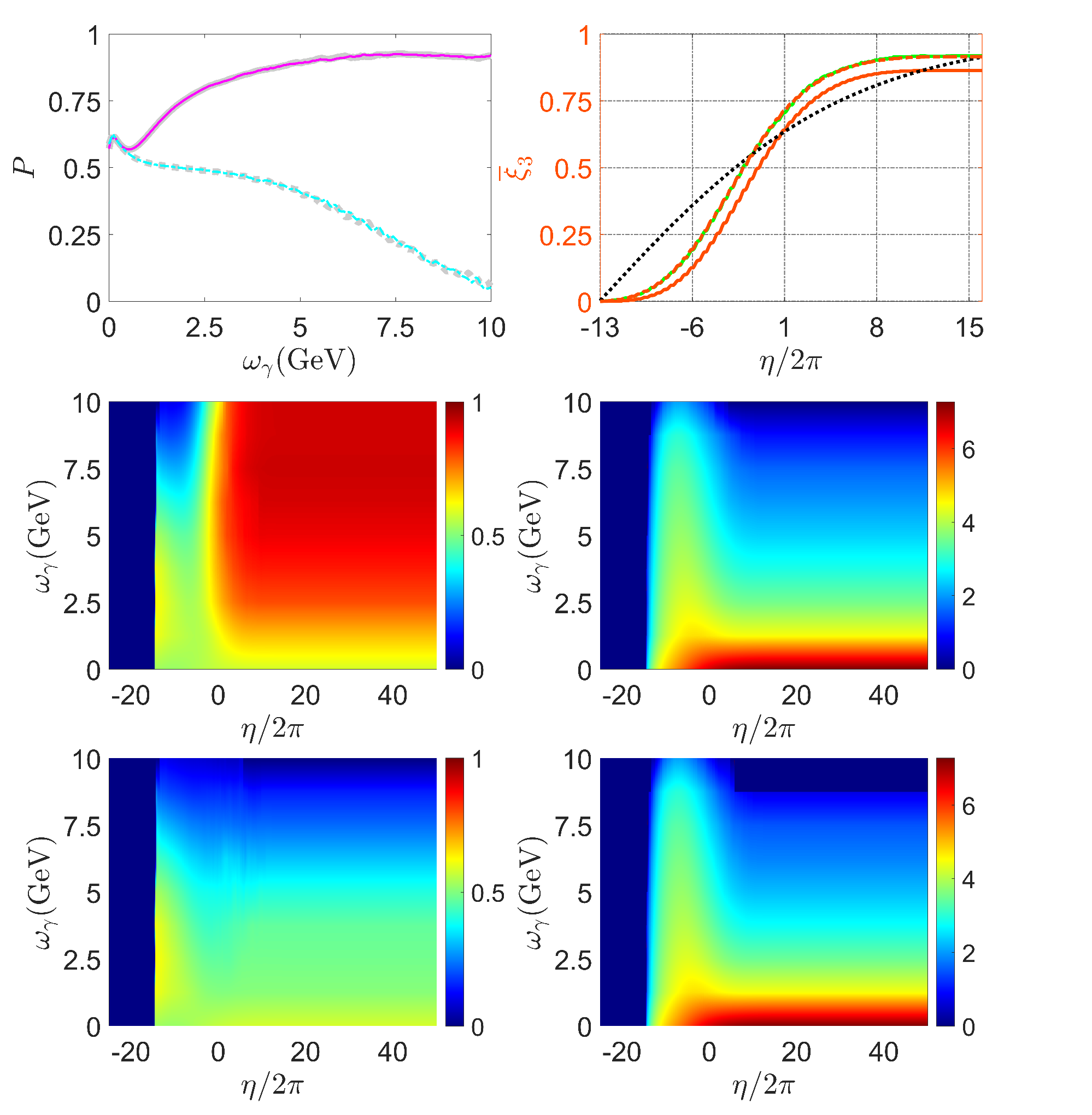}
     \begin{picture}(300,0)   
  \put(25,244){(a)}

  \put(137,244){(b)}
   
     \put(25,162){\color{white}(c)}

  \put(137,162){\color{white}(d)}

   \put(25,81){\color{white}(e)}

  \put(137,81){\color{white}(f)}
\end{picture}	
    \caption{(a): Photon polarization vs $\omega_\gamma$, including only VD (magenta) and VB (cyan) for vacuum polarization. (b): Polarization evolution of a $\gamma$-ray beam [initially $\overline{\bm \xi}=(0,0,0)$] interacting with the laser pulse. Two photon energies, 10 GeV (dash-dotted) and 5 GeV (solid), are considered in the simulations. The green curve corresponds to the result simulated using pure states with half of the photons have $\xi_3=+1$ and half have $\xi_3=-1$. The black-dotted line represents $\xi_3$ calculated by Eq. (\ref{dxi}) with $\omega_\gamma=10$ GeV and a constant $\overline{\chi}_\gamma=4.51$. (c)-(f): Evolutions of polarization (c) and photon number (d) for the electron beam in Fig. \ref{fig2} with VD included, and evolution of polarization (e) and photon number (f) for the same beam with VD excluded, artificially. Here $\eta=\omega_0 t - k_0 z$ is the laser phase in units of $2\pi$.}
    \label{fig3}
\end{figure}

The rationale for achieving high polarization is analyzed in Fig. \ref{fig3}. As demonstrated in Fig. \ref{fig2}(g) that the high photon polarization originates from the vacuum polarization effect, we further investigate the interaction solely with VD or VB, see Fig. \ref{fig3}(a). It becomes evident that only VD leads to high polarization, while the effect of VB is negligible. To provide a clearer understanding of the polarization enhancement by VD, we calculate the polarization evolution of a 10 GeV (5 GeV) $\gamma$-ray beam interacting with the laser pulse. The polarization increases over time to 91.52\% (86.32\%), which aligns with the analytical estimation [the red-dotted line in Fig. \ref{fig3}(b)] derived from Eq. (\ref{xi}), as:
\begin{eqnarray}\label{dxi}
    \frac{d \overline{\xi}_3}{dt}&=&(1-\overline{\xi}_3^2) \int_0^1 dx C_p  {\rm K}_{\frac{2}{3}}(\rho).
\end{eqnarray}
Photons with higher energies experience higher polarization enhancement after an interaction as the $\overline{\xi}_3$ growth rate $\frac{d \overline{\xi}_3}{dt}\sim\int_0^1 dx C_p  {\rm K}_{\frac{2}{3}}(\rho) \propto \chi_\gamma \propto \omega_\gamma$. Therefore, the final photon polarization is proportional to $\omega_\gamma$ in Fig.\ref{fig2}(g). 

The enhancement of polarization due to VD, which involves the differential absorption of photons by the vacuum based on their polarization states, can be intuitively inferred from the second term of Eq. (\ref{Wpair}). It implies that photons with $\xi_3=-1$ are more likely to decay into pairs compared to those with $\xi_3=+1$, resulting in a net polarization. For a more comprehensive understanding, we conduct simulations using pure states in $\xi_3$ axis to describe photon polarization, specifically an unpolarized $\gamma$-ray beam consisting of half $\xi_3=+1$ and half $\xi_3=-1$ photons in Fig. \ref{fig3}(b) (in green). In this time, only the polarization-dependent photon decay (pair production) affects the polarization of the entire beam. As the beam propagates through the laser field, the photon polarization increases over time as more $\xi_3=+1$ photons remain, resulting in a substantial $\overline{\xi}_3=91.52\%$, exactly coincident with the result simulated in mixed states. For a more detailed theoretical explanation from the pure state perspective, please refer to the Supplemental Material \cite{SM}. Note that the polarization variation resulting from asymmetric photon decay is accounted for by Eq. (\ref{xi}) in our simulations using a mixed state approach, which is capable of providing more accurate information in other polarization directions \cite{SM,Liyfei2020}.

\begin{figure}[b]
    \includegraphics[width=1\linewidth]{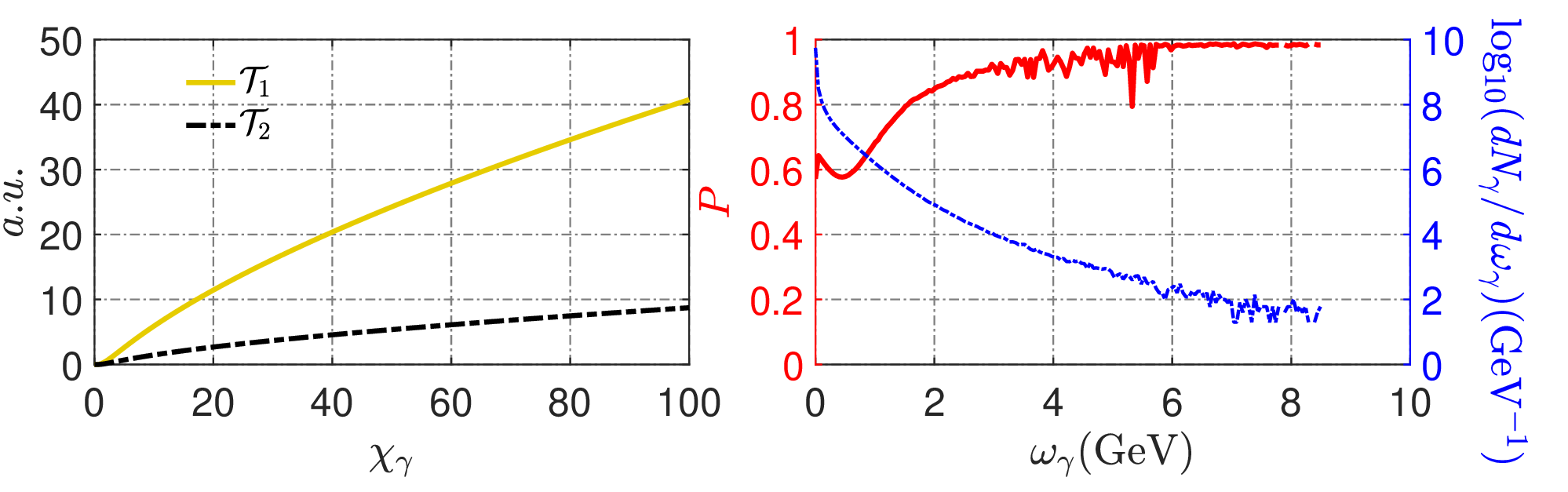}
      \begin{picture}(300,0)   
  \put(16,74){(a)}

  \put(130,74){(b)}
   
\end{picture}	
    \caption{(a): Dependence of $\mathcal{T}_1$ (yellow-solid) and  $\mathcal{T}_2$ (black-dash-dotted) on $\chi_\gamma$, with $\mathcal{T}_1=\int_0^1 dx C_p \left[{\rm IntK}_{\frac{1}{3}}(\rho)+(\frac{x}{1-x}+\frac{1-x}{x}){\rm K}_{\frac{2}{3}}(\rho)\right]$ corresponding to pair-production rate and $\mathcal{T}_2=\int_0^1 dx C_p  {\rm K}_{\frac{2}{3}}(\rho)$ associated with polarization enhancing rate. (b): Photon polarization and number vs $\omega_\gamma$ with $\tau=15 T_0$ and other parameters the same with as in Fig.\ref{fig2}. }
    \label{fig4}
\end{figure}
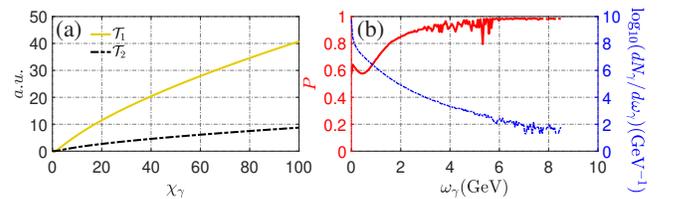

The evolution of polarization and photon number for the electron beam in Fig.\ref{fig2} is illustrated in Figs.\ref{fig3}(c)-(f), with and without the inclusion of VD artificially. Initially, the interaction happens at $\eta/2\pi=-13$, the generated photons exhibit low polarization, as shown in Fig. \ref{fig3}(c). However, over time, the polarization gradually increases, and this increase is proportional to the photon energy. Particularly, for 10-GeV photons, the polarization rises from 0 to 92.7\%. For low-energy photons ranging from 0.5 to 5 GeV, the photon polarization experiences a decline around $\eta/2\pi \approx -10$ due to a sudden increase in photon yield, as shown in Fig. \ref{fig3}(d). In contrast, when VD is excluded, the photon polarization declines over time [see Fig.\ref{fig3}(e)]. This decline occurs because the photons emitted later  have the same $\omega_\gamma$  but encounter a higher $\delta=\omega_\gamma/\varepsilon_e$ compared to the photons emitted earlier due to the decreasing of $\varepsilon_e$ during the interaction. As demonstrated, the decrease in $\xi_3$ is significant with increasing $\delta$ \cite{Liyf2020,SM}, which is also proven by the red-dash-dotted line in Fig.\ref{fig2}(g). The VD induced polarization  also leads to a higher number of $\gamma$-rays, as shown in Figs. \ref{fig3}(d) and (f) [also Fig.\ref{fig2}(g)], owing to a lower pair-production probability with higher $\xi_3$.

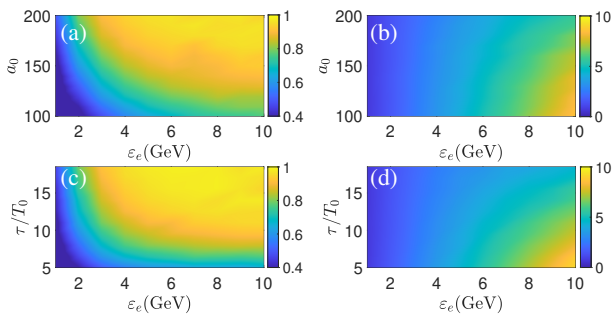
\begin{figure}[tbh]
    \includegraphics[width=1\linewidth]{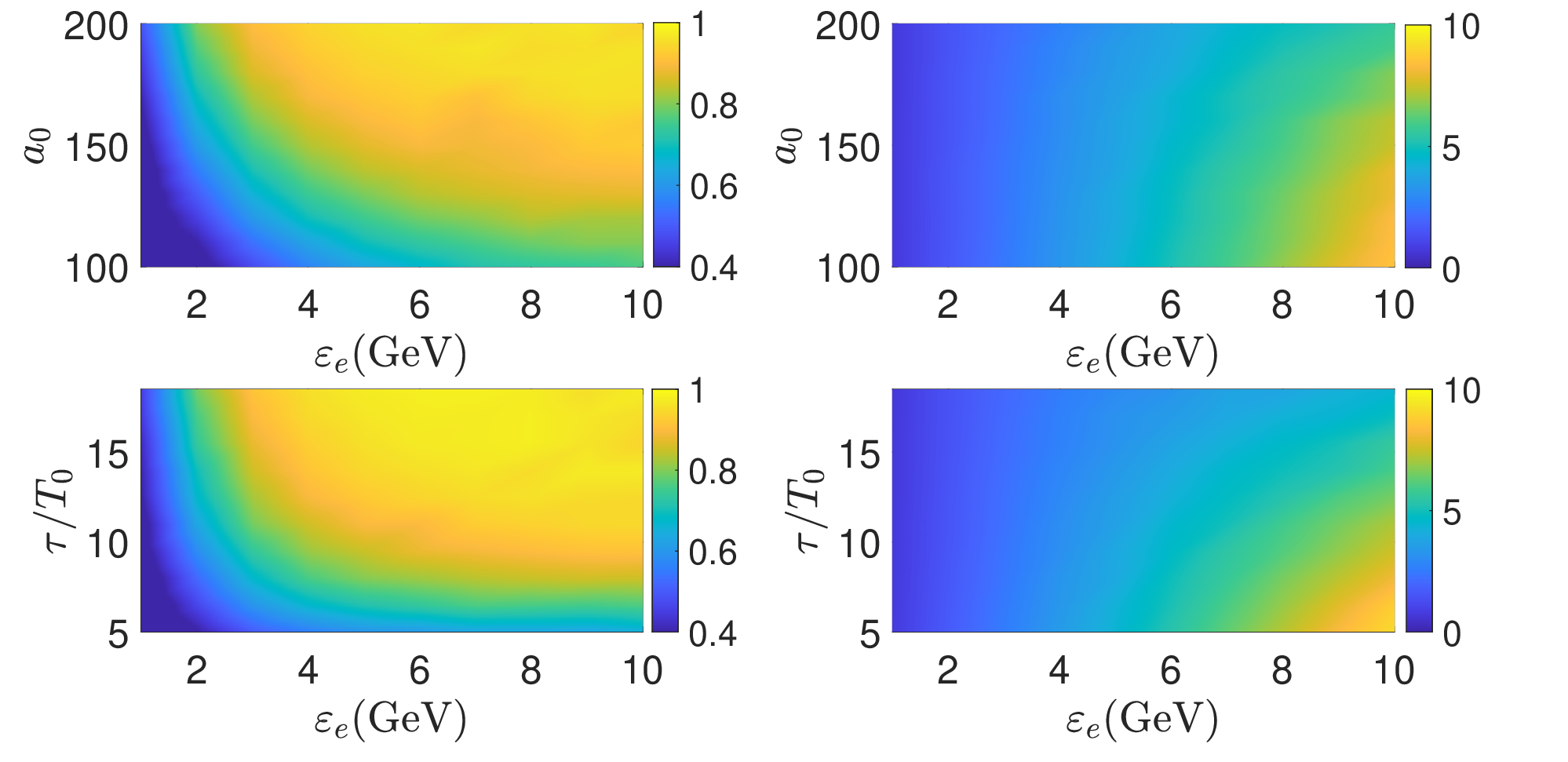}
      \begin{picture}(300,0)   
  \put(24,117){\color{white}(a)}
  \put(141,117){\color{white}(b)}
    \put(24,62){\color{white}(c)}
  \put(141,62){\color{white}(d)}
   
\end{picture}	
    \caption{Distributions of average photon polarization [(a), (c)] and average photon energy [(b), (d)] vs $\varepsilon_e$ and $a_0$ [(a), (b)] or $\tau$ [(c), (b)] for 0.1\% $N_e$ photons selected with highest energies.}
    \label{fig5}
\end{figure}

To assess the practicality of utilizing a highly  polarized, brilliant GeV-$\gamma$-ray beam, we examine the impacts of laser and electron-beam parameters in Fig. \ref{fig4} and Fig. \ref{fig5}. Both the pair-production rate and the polarization enhancing rate  increase with $\chi_\gamma$, see Fig. \ref{fig4}(a). Hence, there exists a trade-off concerning the laser and electron-beam parameters for optimal $\gamma$-ray generation. In addition, the polarization for specific photon energy could be adjusted by changing the laser parameters. For instance, we aim for high polarization at 8 GeV $\leq \omega_\gamma \leq 10$ GeV with $\tau=10$ T$_0$ in Fig. \ref{fig2}. Nonetheless, adjusting $\tau=15$T$_0$ can lead to high polarization of ($98.4\%$) at 6 GeV $\leq \omega_\gamma \leq 8$ GeV, as demonstrated in Fig. \ref{fig4}(b). For a clearer perspective, we present distributions of $\overline{\xi}_3$ and $\overline{\omega}_\gamma$ with respect to  $\varepsilon_e$ and $a_0$ or $\tau$ for 0.1\% $N_e$ photons selected with highest energies in Figs.\ref{fig5}, and show the distributions with respect to $\omega_\gamma$ in \cite{SM}. 

In conclusion, our study delves into the dynamic polarization of photons emitted by an unpolarized electron beam in an ultra-intense laser field, presenting a novel approach to generate a highly polarized, brilliant GeV-$\gamma$-ray beam through an advanced Monte Carlo method. Despite initially low polarization upon photon production, substantial enhancement occurs via vacuum dichroism, which corresponds to asymmetric photon decay between opposite polarization states. Notably, the resulting photon polarization exhibits an energy-dependent pattern, reaching near 100\% at the energy edge. In a feasible all-optical setup featuring a seed electron beam density of $10^8$ per bunch and kinetic energy of 10 GeV, a high-quality $\gamma$-ray beam can be realized with a polarization degree of 92\%, a flux of $\sim 1.46 \times 10^{18}$ phs/s, and an average energy of 6.93 GeV. Moreover, envisioning a possible ultrahigh charge ($\sim$ 100 nC \cite{Ma2018}) for electron beams suggests the potential for an even more brilliant, highly polarized $\gamma$-ray beam, thus advancing the applications of all-optical $\gamma$-ray sources across astrophysics, particle physics, nuclear physics and high-energy physics domains. The resultant polarized $\gamma$-ray beam, with its ultrashort duration, holds promise for potential ultrafast diagnostic applications.\\

We gratefully thank Prof.
Y.-Y. Chen for the helpful discussions.
This work is supported by the
National Natural Science Foundation of China (Grants
No.12222507 and No.12075187), the Strategic Priority
Research Program of the Chinese Academy of Sciences
(Grant No.XDA25031000).

\bibliography{reference}  
\end{document}